\documentclass[floatfix,aps,a4paper,nofootinbib,notitlepage]{revtex4-1}
\usepackage{mathptmx}       
\usepackage{helvet}         
\usepackage{courier}        
\usepackage{makeidx}         
\usepackage{graphicx}        
\usepackage[bottom]{footmisc}
\usepackage[hidelinks,breaklinks=true]{hyperref}
\usepackage{amsopn,dsfont,amssymb}
\usepackage{mathtools}
\mathtoolsset{showonlyrefs=true}
\usepackage{tikz}
\usetikzlibrary{calc,positioning}
\newcommand\textinmath[1]{\text{#1}}

\begin{document}
\title{Bell Correlations and the Common Future}
\author{\"Amin Baumeler}
\author{Julien Degorre}
\author{Stefan Wolf}
\affiliation{Faculty of Informatics, Universit\`{a} della Svizzera italiana, Via G.\ Buffi 13, 6900 Lugano, Switzerland}
\affiliation{Facolt\`{a} indipendente di Gandria, Lunga scala, 6978 Gandria, Switzerland}
\begin{abstract}
	{\em Reichenbach's principle\/} states that in a causal structure, correlations
	of classical information can stem from a common cause in the common
	past or a direct influence from one of the events in correlation to the
	other. The difficulty of explaining {\em Bell correlations\/} through a
	mechanism in that spirit can be read as questioning
	either the principle or even its basis: {\em causality}. In the former case, the principle
	can be replaced by its quantum version, accepting
	as a common cause
	an entangled
	state,  leaving the phenomenon as mysterious as ever on the
	{\em classical\/} level (on which, after all, it occurs). If,  more radically, the
	causal structure is questioned in principle,  closed space-time curves may  become possible that,
	as is argued in the present note, can give rise to  non-local
	 correlations if to-be-correlated pieces of classical information
	meet in the common future --- which they need to  if the correlation is to
	be detected in the first place. The result is a view
	resembling {\em Brassard and Raymond-Robichaud\/}'s parallel-lives variant
	of {\em Hermann's and Everett\/}'s relative-state formalism,
	avoiding   ``multiple realities."
\end{abstract}
\maketitle

\section{Introduction}

Reichenbach's principle~\cite{reichenbach:1956} states that, in a fixed causal structure,  correlations either stem from a {\em common cause\/} or from one part {\em directly influencing\/} the other.
The principle looks natural since it imagines a {\em mechanism\/} that leads to the correlations.
Bell's theorems~\cite{bell,Bell1975} limit the
explanatory power of a {\em common cause\/} in the form of {\em
  classical\/} information while quantum theory predicts correlations
beyond that~--- so-called ``Bell non-local correlations'' (see also~\cite{brown_timpson_2016}). But then, strangely enough,
not all correlations compatible with no-signaling are attainable in nature: An example of an idealization
beyond what quantum physics predicts are  PR correlations~\cite{Popescu1994}, {\em i.e.},
binary inputs~$X,Y$ and  outputs~$A,B$  satisfying~$A\oplus B=XY$.

The absence of a {\em mechanism\/} behind Bell non-local correlations is disturbing, and several patches have been proposed:
One can loosen Reichenbach's principle and simply regard the quantum state as a common cause
(then, no further mechanism is to be expected)~\cite{quantumreichenbach,Allen2016}; one can
resort to one measurement event influencing another (that would have to be an {\em instantaneous fine-tuned\/} influence
in a preferred
frame~\cite{Coretti-Hangi-Wolf-2011,Barnea-Bancal-et-al:2013,Bancal2012,Stefanov-Zbinden-Gisin-Suarez-2002,finetune,Cavalcanti:2017});
one can  assume  signals to travel to the past~\cite{debeauregard1953,Price1994} or suppose the existence of multiple realities~\cite{Brassard2013} | but even at that price, no striking story has been told yet.

If the data in question are never brought  together,  no correlation
can be {\em seen\/}~\cite{Fitzi08}.
At the occasion of that necessary {\em rendez-vous\/} in the future, the
(physically represented) pieces of information  locally interact~| and this
{\em detection procedure\/} of the correlation may be considered its {\em origin\/}:
Such violations of causality can
become possible, {\em e.g.}, through  closed time-like curves~(CTCs).
CTCs are world-lines  {\em closed\/} in time: A system traveling along it can  meet its ``younger self.''
CTCs are {\em consistent\/} with general relativity~\cite{Lanczos:1924,Godel1949}.
We consider the case where all information is {\em classical\/} and all interactions {\em local\/}:
The established correlations are  sent to the past via a CTC as
described by Deutsch~\cite{Deutsch1991},
ending up in a classical story behind Bell non-local correlations.

This text is organized as follows.
 Sections~\ref{sect:Relative causality} and~\ref{sect:From measurement independence to retro-causality}  review specific
modifications of causality, namely Hermann's relative causality and Costa de Beauregard's retro-causality.
  Section~\ref{sect:Retro-causal models}  presents a new formulation of this ``zigzag'' model through relaxing  measurement independence.
    Section~\ref{sect:From relative and retro-causality to closed time-like curves} describes a simple classical mechanism  simulating  Bell non-local correlations.
Section~\ref{Relation to the other models} discusses relations to
previous stories (parallel lives, retro-causality, and  {\em
  Viennese\/} ``process matrices'').

\section{Relative causality \label{sect:Relative causality}}
The works of {\em Grete Hermann\/},  physicist and philosopher, have been strangely overlooked.
Not only did she
spot a mistake (later called ``silly'' by Bell) in von~Neumann's~\cite{VonNeumann1932} ``proof'' that quantum theory cannot be extended to yield deterministic predictions, she also provided her own
arguments against such an extension~| besides contributing to a better understanding of {\em causality}.
In this context, she described the measurement process in the spirit of
the  {\em relative-state interpretation\/}~\cite{Everett1957,EverettThesis} of quantum theory | ten years before {\em Hugh Everett III\/} did.

	Hermann's 1948 article~\cite{Hermann48}  looks into the process of measuring an electron's position as  described quantum-physically:
	If that electron interacts with a photon (described quantum-physically as well), the result is
	\begin{quote}
		        ``a new wave-function which is uniquely determined by the given wave-functions: It does, therefore,  not contain the uncertainty that we would have to attribute to that mystic process.''\footnote{``eine neue Wellenfunktion, die eindeutig durch die beiden gegebenen Wel\-len\-funk\-ti\-onen [\ldots] bestimmt ist. [\ldots] [S]ie enth\"alt also nicht die Unbestimmtheit, die wir jenem mys\-tisch\-en [\ldots] [P]rozess zuschreiben m\"u{\ss}ten.''}
		\end{quote}
		At this stage, one would have to {\em measure\/} this new wave-function to determine the position of the electron.
		\begin{quote}
			        ``Without any such new observations,
                                the quantum-mechanical formalism leads
                                to a progressing not visualizable  braiding of the fundamental particles.''\footnote{``Ohne solche neuen Beobachtungen f\"uhrt der quantenmechanische Formalismus zu einer immer weitergehenden, aber ganz unanschaulichen Verflechtung der Elementarteilchen.''}
			\end{quote}
			According to Hermann, this means that
			\begin{quote}
				        ``the electron, after colliding with the photon, is described by a wave-function with a sharp position only relatively to the new measurement,''\footnote{``Erst relativ zu der neuen Messung wird der Zustand des Elektrons nach seinem Zusammensto{\ss} mit dem Lichtquant durch eine Wellenfunktion mit scharfer Ortsangabe [\ldots] beschrieben [,]''}
				\end{quote}
				and in that perspective (the key of the argument)
				\begin{quote}
					        ``it, therefore,
                                                constitutes an
                                                autonomous physical
                                                system characterized
                                                by its own
                                                wave-function
                                                immediately after the
collision with the photon.''\footnote{``bildet es also unmittelbar nach dem Zusammensto{\ss} mit dem Lichtquant durchaus ein f\"ur sich bestehendes, durch seine eigene Wellenfunktion cha\-rakterisiertes physikalisches System.''}
					\end{quote}
					In other words, after the interaction, the measured system's  wave-function  {\em relates\/} to the measurement outcome~| which is in what she sees the {\em cause\/} for the electron to be at that position.
					Such a cause, however, cannot be brought in to give better predictions as it is only accessible to the experimenter {\em after\/} the measurement.
					In~\cite{HermannFries},  she writes:
					\begin{quote}
						        ``These
                                                        causes could
                                                        not have been
                                                        used for
                                                        predictions;
                                                        they determine
                                                        the system in
                                                        a relative
                                                        way,
                                                        relatively to
                                                        the
                                                        observation
                                                        which was
                                                        obtained only
                                                        at the moment
                                                        of the
                                                        measurement. They,
                                                        therefore,
                                                        could be
                                                        accessed after
                                                        this
                                                        observation
                                                        only and do, hence, not allow to predict the outcome.''\footnote{``Zu einer Voraussage [\ldots] w\"aren jene Gr\"unde [\ldots] nicht zu ge\-brau\-chen; denn auch sie bestimmen [\ldots] das System nur relativ, und zwar re\-lativ zu der Beobachtung, die bei der Messung selber erst gemacht wurde. Sie konnten also dem Physiker erst nach dieser Beobachtung zur Verf\"ugung stehen und ihm somit keine Vorausberechnung von deren Ergebnis gestatten.''}
						\end{quote}
						Through these thoughts, Hermann  anticipates Everett's formalism~\cite{Everett1957,EverettThesis} and, at the same time,  {\em disentangles  causality and predictability}.
More specifically, Hermann~\cite{Hermann1936}  describes an ontology
in which a measurement entangles the observed object to the apparatus,
and only {\em relative  statements\/} are possible.
				Everett goes beyond Hermann's view by
                                invoking the {\em wave-function of the  whole universe}. His formalism
						 has often been called
                                                 ``many worlds:''
                                                 Whenever a system
                                                 gets entangled with the apparatus, all possible results are realized  in {\em parallel\/} universes~--- a~view brought forward, for example, by DeWitt and Deutsch~\cite{DeWitt1970,FabricOfReality}.
						This is a left-over of {\em classical\/} concepts: ``The co\"existing branches [\ldots] can only be related to $\mbox{`worlds'}$ described by classical physics. [\ldots] [T]he [\ldots] meaning of Everett's ideas is not the co\"existence of many [classical] worlds, but on the contrary, the existence of a {\em single quantum one}''~\cite{Levy-Leblond1976}.

						A variation of the Hermann/Everett theme
						are ``{\em parallel
                                                  lives\/}''~\cite{Brassard2013}:
                                                Instead of {\em globally},
                                                the individual
                                                experimenters split {\em
                                                  locally},
                                                into  ``bubbles'' that
                                                are later only visible
to each other  if the quantum predictions result: The
						model is  local {\em
                                                  and\/}  realistic.

	If one incorporates time into the description, then a {\em
          timeless\/} wave-function of the universe  as a whole can be imagined~\cite{Page1983,Wootters1984}:
						The state of {\em one\/} part
                                                of the universe is
                                                determined
                                                  relatively to
                                                {\em another}, called ``clock.''
						By that, all dynamics (the Schr\"odinger equation) can be cast in {\em static form}: {\em Relatively\/} to the clock, the systems undergo the quantum {\em dynamics}.

\section{From measurement-dependence to retro-causality \label{sect:From measurement independence to retro-causality}}
One way of relaxing the causal structure is by a {\em retro-causal effect\/}:
The ``Parisian zigzag" was introduced by Costa de Beauregard seventy years ago~\cite{debeauregard1953}
and recently revived by Price~\cite{Price1994}; we relate it to measurement-dependence.

\subsection{``Parisian zigzag'' \label{sect:Parisian zigzag}}

 In 1947,\footnote{In 1947, Olivier Costa de Beauregard shared this
   idea with Louis de Broglie who disapproved.  It was published in
   1953~\cite{debeauregard-interview}.} Olivier Costa de Beauregard
 questioned  the no-signaling assumption ({\em no instantaneous causality\/}) made by
Einstein, Podolsky, and Rosen ({\em ``EPR''\/} for short) in 1935
  and
considered actions to and from the common past: ``[A]ll the weight of Einstein's argument is moved from instantaneous causality to retroactive causality''\footnote{``[T]out le poids de l'argument d'Einstein est ainsi transport\'e du paradoxe de la causalit\'e imm\'ediate \`a la causalit\'e r\'etroactive".}\cite{debeauregard1953}.
This represents a reply to {\em EPR\/}~--- circumventing the claim to augment quantum theory by hidden
variables --- that can even be seen as a reply to {\em Bell's later reply to  EPR}.

 The ``Parisian zigzag''~\cite{DeBeauregard1976,DeBeauregard1977,Price1994} gives  a
  description of Bell non-local correlations via {\em
    retro-causation}, {\it i.e.}, causation from the future to the
  past.\footnote{In  quantum information,
    Schumacher~\cite{Bennett-super-dense-coding:1992} suggested such a ``zigzag''  for interpreting  superdense coding: ``[O]ne of the two bits is sent forward in time through the treated particle,
while the other bit is sent backward in time to the {\em EPR\/} source, then forward in time through the untreated particle, until finally it is combined with the bit in the treated particle to reconstitute the two-bit message. Because the bit `sent backward in time' cannot be used to
transmit a meaningful message without the help of the other particle, no opportunity for time travel or superluminal communication is created, just as none is created in the classic {\em EPR\/} experiment in which simultaneous measurements are used to establish non-message-bearing correlations over a spacelike interval."}
Assume an experiment in which Alice and Bob each get a photon to be measured.
In that model, the photons ``do not possess polarizations of their own,'' but rather ``borrow one later''~\cite{DeBeauregard1977}:
When Alice performs the measurement on her photon, it gets a random
polarization that is then sent to the
photon's source in the past, from where
the ``borrowed'' polarization travels on to   Bob in the future (this is why the speculation is called {\em ``zigzag''\/}).
In that model, ``Einstein's prohibition to `telegraph to the past' does not hold at the level'' of the photons, but at the one of macroscopic (in other words: classical) objects only~\cite{DeBeauregard1977}.
A crucial point  is that no photon travels {\em directly\/} from one party to the other (a path that is ``physically empty'').
Instead, it goes ``along the Feynman-style zigzag [\ldots] made of two timelike vectors (which is physically occupied).''
The view is related to models~\cite{feldmann95} with  measurement-dependence through a retro-causal effect, perfectly simulating a singlet.
 Section~\ref{sect:Retro-causal models} links  retro-causal approaches to    measurement-dependence.

\subsection{Retro-causal models \label{sect:Retro-causal models}}
The retro-causality of the ``Parisian zigzag''  is related to the relaxation of  measurement independence
in the {\em Bell model\/}:
We denote by $X$ ($Y$) Alice's (Bob's) input, the outputs being $A$ and $B$. The behavior of interest is  a conditional distribution $P_{AB|XY}$.
\begin{figure}
\begin{center}
	\includegraphics[scale=0.6,clip,trim=0cm 0cm 0cm 0cm]{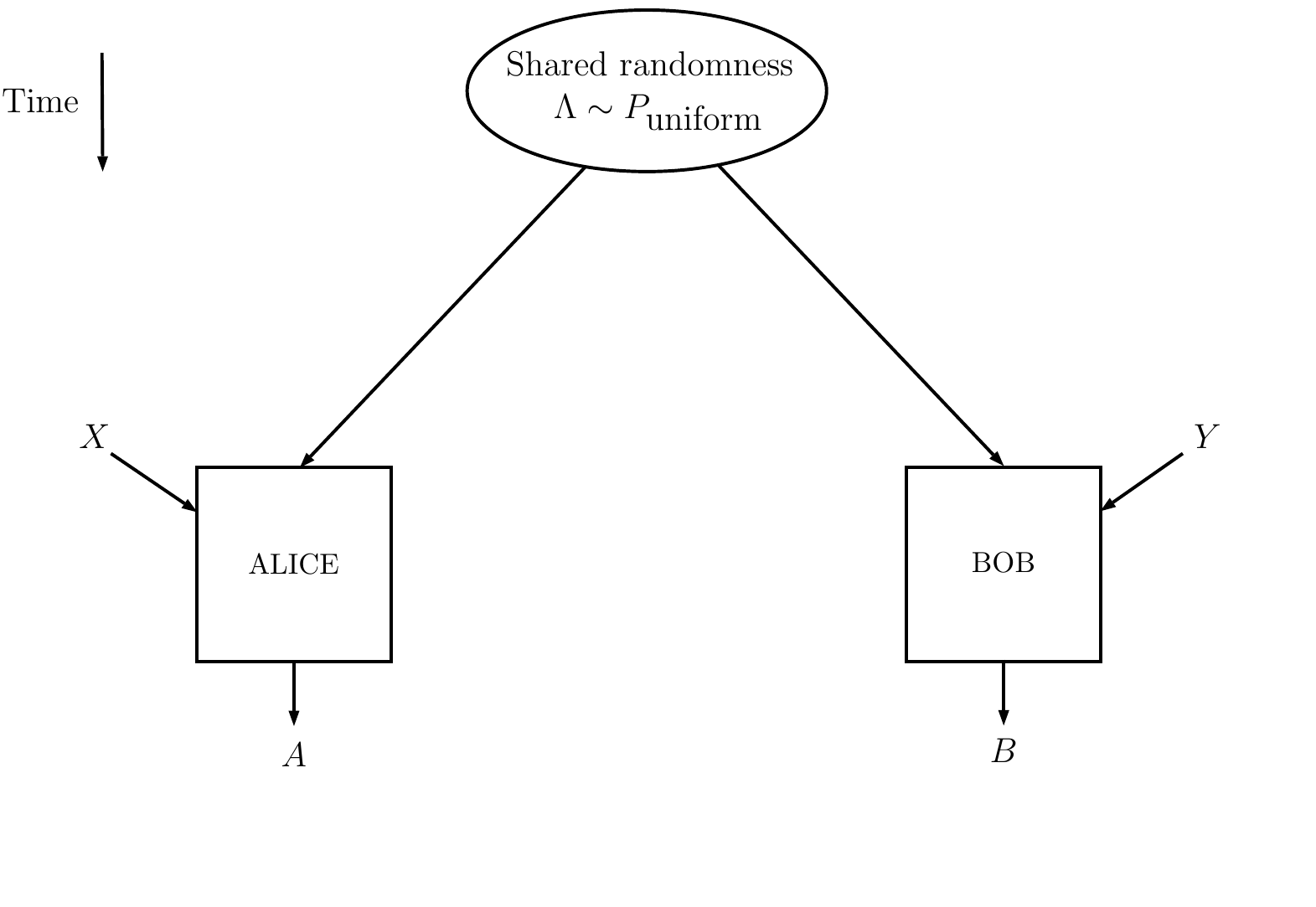}
	\caption{Bell's locality: Time flows from top to bottom.  Alice (Bob) inputs $X$ ($Y$) and obtains $A$ ($B$). Alice and Bob share an infinite amount of randomness~$\Lambda$ distributed independently of Alice's and Bob's inputs. }
	\label{fig:LHV}
\end{center}
\end{figure}
Bell-locality allows Alice and Bob to share an infinite amount of randomness~$\Lambda$ (see Figure~\ref{fig:LHV}):
 A  distribution $P_{AB|XY}$ is called \mbox{\emph{Bell-local}} if it can be written as $$P_{AB|XY}=\sum_{\lambda}P_{\Lambda}(\lambda)P_{A|X,\Lambda=\lambda}P_{B|Y,\Lambda=\lambda}\ .$$
It is remarkable that quantum theory is consistent with  correlations which are {\em not\/} Bell-local~\cite{bell}.
The definition of  {\em Bell-locality\/} decomposes into
three conditions.
\begin{enumerate}
\item \emph{No-signaling}: Alice's output is independent of Bob's input, and {\em vice versa}.
  \item \emph{Locality}: The correlations between Alice and Bob stem from a shared  random variable $\Lambda$.
  \item \emph{Measurement independence}: The shared randomness is independent of Alice's and Bob's inputs.
\end{enumerate}
The third assumption is usually  implicit as $\Lambda$ is understood to root in the {\em common past\/} of Alice and Bob.
If we allow the shared randomness to depend on Alice's and Bob's inputs, it is possible to reproduce the  joint  distribution of a Bell non-local quantum state~\cite{Brans1988}.
Relaxing  this assumption can mean, {\em e.g.}, that the shared
randomness  influences the measurement settings of Alice and Bob. This
does not change the causal structure: The common cause is in the past,
and there is no retro-causal effect. (This is a flavor of determinism
or the fully causal hidden variable approach of
Brans~\cite{Brans1988,Hall2016}).
Alternatively,
it can mean that
the inputs influence the shared
randomness. This points to the distributed-sampling
problem~\cite{Degorre2005} and   retro-causal models as discussed above.

\begin{figure}
\begin{center}

	\includegraphics[clip,trim=0cm 0cm 0cm 0cm,width=10cm]{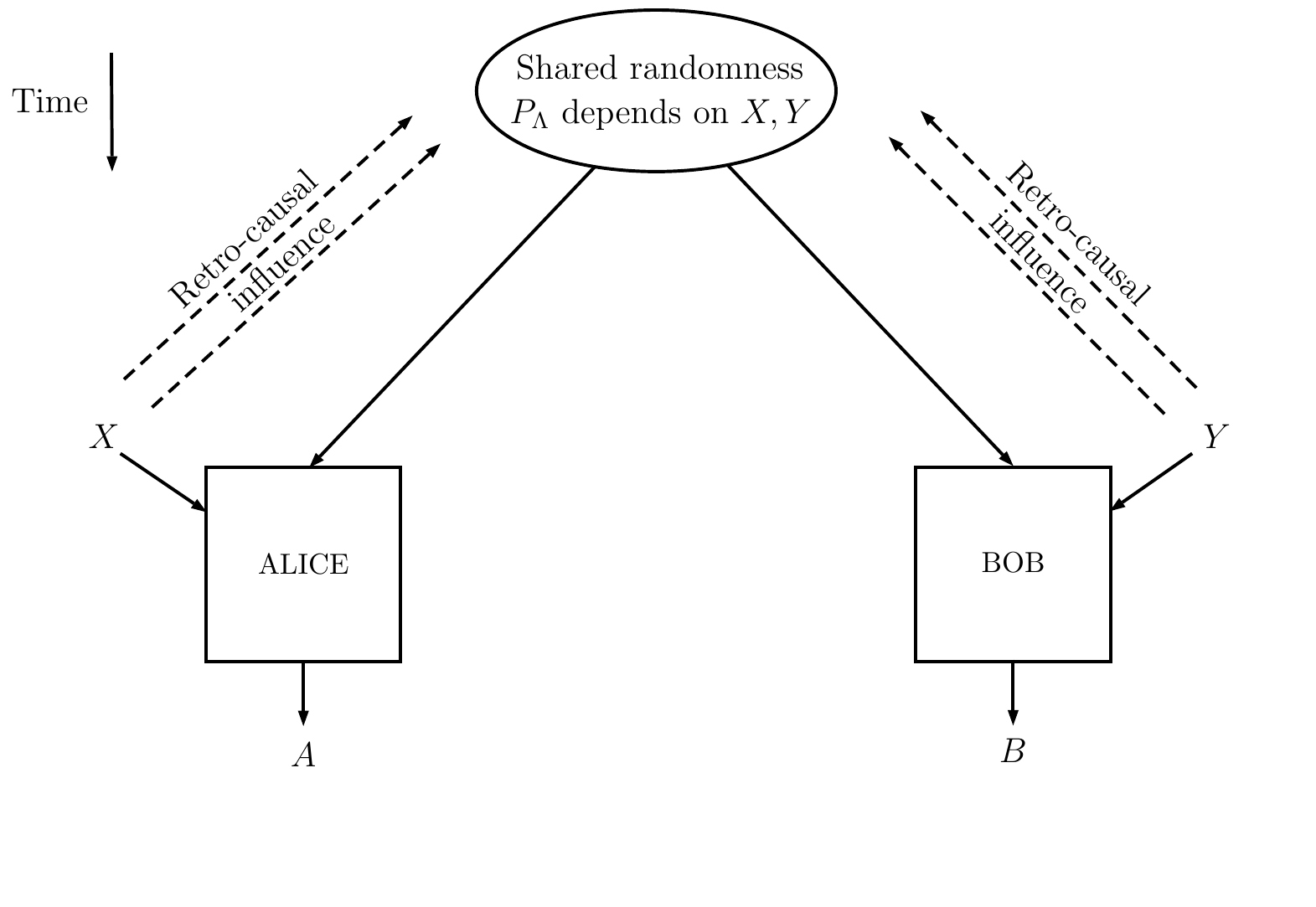}
	\caption{A Bell-like model with  measurement-dependence and   retro-causal influence: The  distribution of the shared random variables depends on  Alice's and Bob's inputs.}
	\label{fig:LHV-retro}
\end{center}
\end{figure}

Here, we are  interested in a model with relaxed  causal structure and
 focus on the second option: A Bell-local model with measurement
 dependence  through a  \emph{retro-causal influence} (Figure~\ref{fig:LHV-retro}).
This  model can reproduce  no-signaling correlations:
A~first idea is to send  Alice's input   retro-actively  to the common past and to share it  with Bob.
Alternatively, Alice's input can be retro-causally used to bias the uniform  distribution of the shared randomness ($\Lambda \sim P_{{\rm uniform}}$).
This is the retro-causal model of Feldman \cite{feldmann95},
solving a \emph{distributed-sampling problem}, and it has been shown
to allow Alice and Bob for  simulating a singlet~\cite{Degorre2005}:
The Toner-Bacon protocol with one bit of communication translates to a ``zigzag'' with a retro-causal bit (from Alice to the common past).
Note that in these models, the mutual information between the shared randomness  $\Lambda$ and  Alice's input is nonzero.\footnote{There are numerous results on calculating and minimizing its amount~\cite{Toner:2003dp,Kofler-Paterek-Brukner:2006,Roland2009,Gisin-Barett:2011,Hall2016}.} It means that there is a  \emph{hidden influence\/}  going from Alice to Bob via a fine-tuned ``zigzag''~\cite{finetune}.
Alternatively, we can build a mechanism
{\em without signaling\/} between Alice and Bob.
 We use the fact~\cite{Degorre2005}   that the protocol
 reproducing a maximally entangled state with the help of one PR-box~\cite{cgmp05} leads  to a ``zigzag-style protocol''
with two ``retro-causal bits:'' One~bit each   travels from Alice and
Bob, respectively, to their common past, the PR~correlation is
established there, and  the
singlet can be simulated.

\section{From relative and retro-causality to closed time-like curves\label{sect:From relative and retro-causality to closed time-like curves}}
Both  models
of Section~\ref{sect:Retro-causal models}
have in common that the future affects  the past.
In  ``parallel lives,'' this manifests itself in the parties  {\em
  meeting-up\/} for the correlation to be established/detected.
In the ``zigzag''  models, the inputs of Alice and Bob
are sent to the common past and the correlation is established there.
A~combination of these pictures sees
the respective data meet in the {\em future},
and the local computation necessary for the {\em verification\/} of the correlation
is at the same time {\em its origin\/}~| if the data can travel back in time via a {\em closed time-like curve
  (CTC)}.\footnote{The latter
have already  been widely discussed in  quantum  information~\cite{Deutsch1991,Aaronson:2009,Bacon-ctc:2004,Mile2015}.}

\subsection{Closed time-like curves with classical information}
\begin{figure}
	\centering
	\begin{tikzpicture}
		\node[draw,rectangle,minimum width=1cm,minimum height=1cm] (U) {$\varepsilon$};
		\draw[<-] (U.250) -- ++(0,-1) node[below] {$P^\text{init}_R$};
		\draw[->] (U.110) -- ++(0,1) node[above] {$P^\text{fin}_R$};
		\draw[->,rounded corners=0.2cm] (U.70) -- ++(0,0.5) -- ++(0.7,0) -- ++(0,-2) node[midway,right] {$P^\text{cons}_V$} -- ++(-0.7,0) -- (U.290);
	\end{tikzpicture}
	\caption{A causality-respecting system jointly
          evolves with a causality-violating one: The latter
          system's state is the fixed-point with maximal entropy.}
	\label{fig:dctc}
\end{figure}
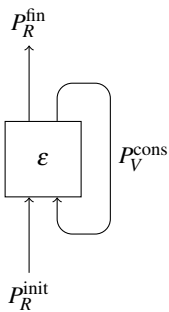

The  idea of Deutsch's model for CTC dynamics is that two systems
undergo a joint evolution after which one of them travels back to the past and re-enters. Whereas
Deutsch described his model for {\em quantum\/} states,
we use {\em classical\/} information~\cite{Aaronson:2009}.
The {\em causality-respecting\/} system is denoted by~$R$,  the {\em causality-violating\/} one by~$V$, and
the joint evolution is~$\varepsilon=P_{R'V'\mid RV}$.
For an  initial state~$P^\text{init}_R$ of~$R$ and given~$\varepsilon$, Deutsch's consistency condition is
\begin{align}
	P^\text{cons}_V = \sum_{r',r,v}P_{R'=r',V'\mid R=r,V=v} P^\text{init}_{R=r} P^\text{cons}_{V=v}
	\,,
\end{align}
{\em i.e.}, the states of~$V$ {\em before\/} and~{\em after\/} the evolution are identical (see Figure~\ref{fig:dctc}).
Generally, {\em several\/} consistent states can exist:
In that case, Deutsch suggests to choose the one  maximizing the entropy, avoiding the {\em information antinomy}.
The final state~$P^\text{fin}_R$ of the causality-respecting system is then
\begin{align}
	P^\text{fin}_R = \sum_{v',r,v'} P_{R',V'=v'\mid R=r,V=v} P^\text{init}_{R=r} P^\text{cons}_{V=v}
	\,.
\end{align}

\subsection{No-signaling correlations from  closed time-like curves}
\begin{figure}
	\centering
\newcommand{\mouth}[2]{
		\foreach\i in {0,0.1,...,1}{
			\draw[fill=black,opacity=1-\i] #1 ellipse ({\i*0.4cm} and {\i*0.2cm}) node (#2) {};
		}
}
	\begin{tikzpicture}
		\node[draw,rectangle,minimum width=1.5cm,minimum height=1cm] (A) {};
		\node[draw,rectangle,minimum width=1.5cm,minimum height=1cm,right=2cm of A] (B) {};
		\draw[->] (A.north) to[out= 90,in=280] ++(-0.25, 0.5) node[right] {$A$};
		\draw[->] (B.north) to[out= 90,in=260] ++( 0.25, 0.5) node[left]  {$B$};
		\draw[<-] (A.south) to[out=270,in= 80] ++(-0.25,-0.5) node[right] {$X$};
		\draw[<-] (B.south) to[out=270,in=100] ++( 0.25,-0.5) node[left]  {$Y$};
		\draw[-,dashed,rounded corners=1cm] ($ (A.north west) - (1.5, 3) $) -- ($ (A.north) + (0,1.5) $) -- ($ (A.north east) - (-1.5, 3) $);
		\draw[-,dashed,rounded corners=1cm] ($ (B.north west) - (1.5, 3) $) -- ($ (B.north) + (0,1.5) $) -- ($ (B.north east) - (-1.5, 3) $);
		\draw[-,dashed,rounded corners=1cm] ($ (A.south west) - (2,-4) $) -- ($ (A.south) - (0,1.5) $) -- ($ (A.south east) - (-2,-4) $);
		\draw[-,dashed,rounded corners=1cm] ($ (B.south west) - (2,-4) $) -- ($ (B.south) - (0,1.5) $) -- ($ (B.south east) - (-2,-4) $);
		\mouth{($ (A)!0.2!(B) +(0,-2) $)}{AP}
		\mouth{($ (A)!0.8!(B) +(0,-2) $)}{BP}
		\draw[->] (AP) to[out=120,in=300] node[midway,left] {$\Lambda_1$} (A) coordinate (Ain);
		\draw[->] (BP) to[out= 60,in=240] node[midway,right] {$\Lambda_2$} (B) coordinate (Bin);
		\draw[-,dashdotted] (Ain) -- (A.north);
		\draw[-,dashdotted] (Bin) -- (B.north);
		\mouth{($ (A)!0.5!(B) +(0,3.5) $)}{Lfuture}
		\node[draw,rectangle,minimum width=0cm,minimum height=0cm] at ($ (A)!0.5!(B) +(0,+2.5) $)  (PR) {$\mathsf{PR}$};
		\draw[->] (A) to[out=60,in=250] node[midway,left] {$M_1$} (PR);
		\draw[->] (B) to[out=120,in=290] node[midway,right] {$M_2$} (PR);
		\draw[->] ($ (PR.north) + (-0.05,0) $) -- ++(0,0.6) node[midway,left] {$\Lambda'_1$};
		\draw[->] ($ (PR.north) + (+0.05,0) $) -- ++(0,0.6) node[midway,right] {$\Lambda'_2$};
		\draw[-,dashdotted] (A.south) -- ($ (A.north) + (0.3,0) $);
		\draw[-,dashdotted] (B.south) -- ($ (B.north) + (-0.3,0) $);
	\end{tikzpicture}
	\caption{A PR~box is simulated locally in the common future of Alice and Bob. The outputs of the PR~box  travel back in time trough a Deutsch closed time-like curve. The dashed lines represent the light cones. Time flows bottom-up.}
	\label{fig:dctcpr}
\end{figure}
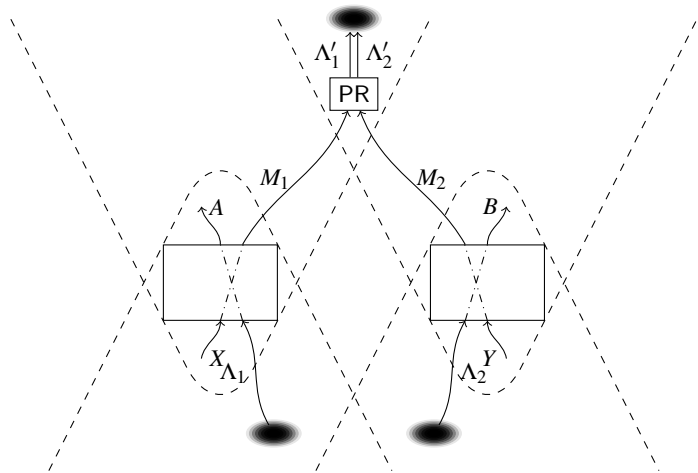

We present the setup with Deutsch CTCs based on random variables to
reproduce any  no-signaling correlation.
We show the representative example of
the {\em Popescu/Rohrlich (PR) box\/}~\cite{Popescu1994}  defined as (see Figure~\ref{fig:PR}):
\begin{align}
	\label{eq:PR}
	P^\textinmath{PR}_{AB\mid XY}(a,b,x,z) = \frac{\delta_{xy,a\oplus b}}{2}
	\,.
\end{align}
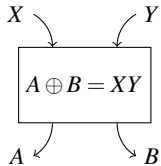
\begin{figure}
	\centering
	\begin{tikzpicture}
		\node[draw,rectangle,minimum width=1.619cm,minimum height=1cm] (PR) {$A\oplus B=XY$};
		\draw[<-] (PR.50) arc (180:120:0.5) node[right] {$Y$};
		\draw[<-] (PR.130) arc (0:60:0.5) node[left] {$X$};
		\draw[->] (PR.230) arc (0:-60:0.5) node[left] {$A$};
		\draw[->] (PR.310) arc (180:240:0.5) node[right] {$B$};
	\end{tikzpicture}
	\caption{A PR box with binary inputs ($X,Y$) and outputs ($A,B$) satisfying~$A\oplus B=XY$.}
	\label{fig:PR}
\end{figure}
 If Alice and Bob  want to simulate the PR box   with
  {\em shared classical randomness}, they can reach a success
  probability of~$3/4$.
When, instead, Alice and Bob share a quantum state, then (at
most~\cite{TsirelsonsBound}) roughly~85\% is possible.
If, on the other hand, Alice and Bob have access to a classical CTC,
they can {\em perfectly\/} simulate a PR box  by {\em local\/} interactions.
The idea is that Alice and Bob send their inputs~$X,Y$ to the common
future
to have them  interact  {\em locally},
resulting in outputs  according to the PR condition, and let them
 travel back along  the CTC (see Figure~\ref{fig:dctcpr}).

The  setup  uses an ``open'' time-like curve~\cite{Pienaar2013,Mile2015}: The systems  traveling to the past do {\em not\/} self-interact.
When the local operations are swaps,  then a {\em single\/} fixed
point of the evolution exists (the PR box); this avoids the {\em information antinomy}.
Note that the setup does not become signaling even
if Alice or Bob choose to apply a {\em different\/} operation locally.
First, the joint  distribution $P^\textinmath{PR}_{AB\mid
  XY}(a,b,x,z)$ of the PR box is {\em no-signaling\/} and second, the
state of all systems {\em just before\/} the parties apply their local
operations is~$P_XP_Y\rho$, where Alice's part of $\rho$ contains no
information on $Y$, and {\em vice versa\/}:
Deutsch CTCs are  {\em no-signaling preserving}.
(They are, however, still  an ``overkill'' because they allow for reproducing {\em any\/} no-signaling  distribution.
A question worth exploring is to find a consistent mechanism, weaker than classical Deutsch CTCs, restricting
the resulting correlations~| ideally to exactly the {\em quantum\/} correlations.)

\section{Relation to the other models \label{Relation to the other models}}

We
discuss the relation of
our speculation
to
previously considered ``stories'' behind the emergence of Bell correlations.

\subsection{Hermann/Everett  and ``parallel lives''}
The parallel-lives model~\cite{Brassard2013} (see also~\cite{Deutsch-Hayden:2000}) assumes that every party, when performing a measurement, splits into ``bubbles'' in different realities,
labeled by the measurement outcome.
When Alice and Bob meet in the common future to compare their results, only those ``bubbles''
are visible to each other for which the labels  reproduce the desired  correlation.
CTCs are an alternative to  such a matching rule.

\subsection{Retro-causality and distributed sampling}
In the retro-causal ``Parisian zigzag''~\cite{debeauregard1953,DeBeauregard1976,DeBeauregard1977,Price1994,price2015}, the input of Alice is sent to the past where it influences the shared random variable of Bob. According to  Section~\ref{sect:Retro-causal models}, this model is {\em fine-tuned}~\cite{finetune} and harmonizes with our own speculation in the sense that
a CTC
can be seen as a mechanism for achieving  retro-causality
operationally and in a consistent way, {\em i.e.}, without time-travel antinomies.

\subsection{The ``process-matrix'' framework}
In~\cite{OCB}, no causal structure is assumed {\it a priori\/} but only
 local assumptions  are made: The
parties receive a system from the environment, interact with it, and
output it back to the environment.
The latter is non-causal: The inputs to the parties can depend on their outputs.
The framework allows for  correlations incompatible with definite orders of the parties.
 Since  such correlations  can be obtained even for the classical
 case~\cite{Baumeler2016}, one might wonder whether also Bell-like
 correlations can.
This is, however, not so.
More specifically, the question is whether Bell non-local correlations
can be seen as 
arising from some classical (as opposed to quantum) mechanism which is
non-causal. 
The answer to this question is: ``If yes, then only at the expense of
signaling.'' 
Thus, the classical variant of the process-matrix framework does not
allow for a 
{\em non-signaling\/} explanation of Bell non-local correlations. 
The reasoning is straight-forward: Non-signaling correlations are
causal 
(they can be simulated in a causal way); and the contra-positive
thereof 
is that non-causal correlations are signaling. In other words, any
classical process 
matrix that allows for non-causal correlations --- this is the 
surplus of the process-matrix framework --- allows for signaling as well.

\section{A look back and a look forward}

 In Sections~\ref{sect:Relative causality} and~\ref{sect:From
   measurement independence to retro-causality}, we revisit two
 routes~--- proposed more than
twenty years {\em before\/} Bell's argument ---  to relaxing the causal structure for  reconciling  Reichenbach's principle and quantum correlations: Hermann's relative causality and Costa de Beauregard's retro-causality.
  In Section~\ref{sect:Retro-causal models}, we show a new formulation
  of the ``zigzag'' model rooted in measurement-dependence
 and shedding light on  the hidden signaling involved.
In Section~\ref{sect:From relative and retro-causality to closed
  time-like curves}, we realize the retro-causal effect
with {\em closed time-like curves}, hereby speculating
about
 a
{\em classical\/} mechanism establishing  {\em Bell non-local correlations}.

When the parts of a system in an entangled quantum state are measured,
then
shared classical information can be insufficient for
explaining the observed correlations: {\em John Stewart Bell's ``exploit''\/}  in  1964
questioned fundamentally
the validity of an
attack to quantum theory by Einstein, Podolsky, and Rosen in
  1935~---
but the puzzle is, after Bell, as unsolved as ever:
What could be a classical {\em mechanism\/} leading to correlations of classical  information?
Reichenbach's principle states that in a given causal structure,
this
can root either
in a {\em common cause\/} in the common past or in a {\em direct influence\/} from one of the correlated events to the other.
Various results question its applicability in the light of Bell correlations~| not only but in particular in the
{\em multipartite\/} scenario~\cite{Stefanov-Zbinden-Gisin-Suarez-2002,finetune,Coretti-Hangi-Wolf-2011,Barnea-Bancal-et-al:2013,Bancal2012}.
 There are at least three escapes from the  dilemma: First, Reichenbach's principle
is declared wrong,  Bell correlations being a counterexample. It could
then,
second, be
replaced  by a modified~| {\em quantum\/}~| principle accepting as a reason for
correlations of classical pieces of information also  an  {\em entangled state}. Third, we can drop the assumption of  a fundamental
causal structure (Reichenbach's principle's basis).
With the first two ``{\em emergency exits},'' the story ends  here;
we consider the third option: Drop fundamental
causality.\footnote{Are we throwing out the baby with the bath water?
  Maybe~| at least, this has some tradition: ``The law of causality [\ldots] is a relic of a bygone age, surviving, like
the monarchy, only because it is erroneously supposed to do no harm." (Bertrand
Russell, 1913.)}
This text is concerned with how  {\em loopholes\/} of
rigid causality, such as closed timelike curves, can be used to obtain
Bell violations.
Let us finish with a wilder speculation: {\em What if,
in the spirit of Wheeler's   ``It from Bit"~\cite{Wheeler:1989},
space-time causality  emerges from
``laws of large numbers'' at the macroscopic level
of the
thermodynamic limit  hand in hand  with~|  and not   prior to~| the
classical\/}\footnote{Classicality is an  idealized  notion existing
in the  thermodynamic
limit and
related to {\em macroscopicity and
redundancy\/} ({\em i.e.}, work value~\cite{Baumeler-Wolf-ccc:2016}). Violations of Bell's
inequalities indicate
the emergence, in a space-like separated way,  of identical
  classical bits. {\em This\/} is the
strangeness about the missing classical story:
The   {\em same\/} decision seems to be taken  {\em twice\/}!

The measurement is the transition from quantum to {\em classical\/}
information, so one may look
for the key to the quantum measurement problem~| and now also: the
emergence of spacetime
causality when seen as a feature of the classical realm~|
 within {\em thermodynamics\/}
({\em combinatorics\/}), well aware of the fact that none of the
present {\em interpretations of quantum theory\/} harmonizes with Bell
non-locality.
(Surely, the correlations are sometimes used by camp~$X$ to question camp~$Y$'s interpretation and {\em vice versa\/}~|  but this often pairs with   blindness for
the weakness
of one's own favorite metaphysics
in the face of the {\em Bell curse\/}; or as
 {\em Nicolas Gisin\/} put it:
{\em ``Bell, c'est difficile pour tout le monde."})}
{\em information so strangely correlated? (Can we come up with a coherent
  combinatorial
canvas
comprehending
 the creation of classicality,
  correlations, and causality~| combined?)\/}

\begin{acknowledgments}
We thank Andrei Khrennikov, Nicolas Gisin, and Nicolas Brunner for
their kind invitation to the fascinating event in V\"axj\"o.
This research is supported by the Swiss National Science Foundation (SNF), the National Centre of Competence in Research “Quantum Science and Technology” (QSIT), the COST action on Fundamental Problems in Quantum Physics, and by the Hasler Foundation.
\end{acknowledgments}

\bibliography{ref-springer}

\begin{thebibliography}{55}%
\makeatletter
\providecommand \@ifxundefined [1]{%
 \@ifx{#1\undefined}
}%
\providecommand \@ifnum [1]{%
 \ifnum #1\expandafter \@firstoftwo
 \else \expandafter \@secondoftwo
 \fi
}%
\providecommand \@ifx [1]{%
 \ifx #1\expandafter \@firstoftwo
 \else \expandafter \@secondoftwo
 \fi
}%
\providecommand \natexlab [1]{#1}%
\providecommand \enquote  [1]{``#1''}%
\providecommand \bibnamefont  [1]{#1}%
\providecommand \bibfnamefont [1]{#1}%
\providecommand \citenamefont [1]{#1}%
\providecommand \href@noop [0]{\@secondoftwo}%
\providecommand \href [0]{\begingroup \@sanitize@url \@href}%
\providecommand \@href[1]{\@@startlink{#1}\@@href}%
\providecommand \@@href[1]{\endgroup#1\@@endlink}%
\providecommand \@sanitize@url [0]{\catcode `\\12\catcode `\$12\catcode
  `\&12\catcode `\#12\catcode `\^12\catcode `\_12\catcode `\%12\relax}%
\providecommand \@@startlink[1]{}%
\providecommand \@@endlink[0]{}%
\providecommand \url  [0]{\begingroup\@sanitize@url \@url }%
\providecommand \@url [1]{\endgroup\@href {#1}{\urlprefix }}%
\providecommand \urlprefix  [0]{URL }%
\providecommand \Eprint [0]{\href }%
\providecommand \doibase [0]{http://dx.doi.org/}%
\providecommand \selectlanguage [0]{\@gobble}%
\providecommand \bibinfo  [0]{\@secondoftwo}%
\providecommand \bibfield  [0]{\@secondoftwo}%
\providecommand \translation [1]{[#1]}%
\providecommand \BibitemOpen [0]{}%
\providecommand \bibitemStop [0]{}%
\providecommand \bibitemNoStop [0]{.\EOS\space}%
\providecommand \EOS [0]{\spacefactor3000\relax}%
\providecommand \BibitemShut  [1]{\csname bibitem#1\endcsname}%
\let\auto@bib@innerbib\@empty
\bibitem [{\citenamefont {Reichenbach}(1956)}]{reichenbach:1956}%
  \BibitemOpen
  \bibfield  {author} {\bibinfo {author} {\bibfnamefont {H.}~\bibnamefont
  {Reichenbach}},\ }in\ \href@noop {} {\emph {\bibinfo {booktitle} {The
  Direction of Time}}},\ \bibinfo {editor} {edited by\ \bibinfo {editor}
  {\bibfnamefont {M.}~\bibnamefont {Reichenbach}}}\ (\bibinfo  {publisher}
  {University of California Press},\ \bibinfo {address} {Berkeley},\ \bibinfo
  {year} {1956})\ Chap.~\bibinfo {chapter} {19}, pp.\ \bibinfo {pages}
  {157--167}\BibitemShut {NoStop}%
\bibitem [{\citenamefont {Bell}(1964)}]{bell}%
  \BibitemOpen
  \bibfield  {author} {\bibinfo {author} {\bibfnamefont {J.~S.}\ \bibnamefont
  {Bell}},\ }\href@noop {} {\bibfield  {journal} {\bibinfo  {journal}
  {Physics}\ }\textbf {\bibinfo {volume} {1}},\ \bibinfo {pages} {195}
  (\bibinfo {year} {1964})}\BibitemShut {NoStop}%
\bibitem [{\citenamefont {Bell}(1975)}]{Bell1975}%
  \BibitemOpen
  \bibfield  {author} {\bibinfo {author} {\bibfnamefont {J.~S.}\ \bibnamefont
  {Bell}},\ }\href {\doibase 10.1111/j.1746-8361.1985.tb01249.x} {\emph
  {\bibinfo {title} {The theory of local beables}}},\ \bibinfo {type} {Tech.
  Rep.}\ (\bibinfo  {institution} {CERN},\ \bibinfo {year} {1975})\ \bibinfo
  {note} {presented at the sixth GIFT Seminar, Jaca, 2--7 June 1975, and
  reproduced in {\it Epistemological Letters}, March 1976, and in {\it
  dialectica}, June 1985.}\BibitemShut {Stop}%
\bibitem [{\citenamefont {Brown}\ and\ \citenamefont
  {Timpson}(2016)}]{brown_timpson_2016}%
  \BibitemOpen
  \bibfield  {author} {\bibinfo {author} {\bibfnamefont {H.~R.}\ \bibnamefont
  {Brown}}\ and\ \bibinfo {author} {\bibfnamefont {C.~G.}\ \bibnamefont
  {Timpson}},\ }in\ \href {\doibase 10.1017/CBO9781316219393.008} {\emph
  {\bibinfo {booktitle} {Quantum Nonlocality and Reality: 50 Years of Bell's
  Theorem}}},\ \bibinfo {editor} {edited by\ \bibinfo {editor} {\bibfnamefont
  {M.}~\bibnamefont {Bell}}\ and\ \bibinfo {editor} {\bibfnamefont {S.~E.}\
  \bibnamefont {Gao}}}\ (\bibinfo  {publisher} {Cambridge University Press},\
  \bibinfo {address} {Cambridge},\ \bibinfo {year} {2016})\ p.\ \bibinfo
  {pages} {91–123}\BibitemShut {NoStop}%
\bibitem [{\citenamefont {Popescu}\ and\ \citenamefont
  {Rohrlich}(1994)}]{Popescu1994}%
  \BibitemOpen
  \bibfield  {author} {\bibinfo {author} {\bibfnamefont {S.}~\bibnamefont
  {Popescu}}\ and\ \bibinfo {author} {\bibfnamefont {D.}~\bibnamefont
  {Rohrlich}},\ }\href {\doibase 10.1007/BF02058098} {\bibfield  {journal}
  {\bibinfo  {journal} {Foundations of Physics}\ }\textbf {\bibinfo {volume}
  {24}},\ \bibinfo {pages} {379} (\bibinfo {year} {1994})}\BibitemShut
  {NoStop}%
\bibitem [{\citenamefont {Cavalcanti}\ and\ \citenamefont
  {Lal}(2014)}]{quantumreichenbach}%
  \BibitemOpen
  \bibfield  {author} {\bibinfo {author} {\bibfnamefont {E.~G.}\ \bibnamefont
  {Cavalcanti}}\ and\ \bibinfo {author} {\bibfnamefont {R.}~\bibnamefont
  {Lal}},\ }\href@noop {} {\bibfield  {journal} {\bibinfo  {journal} {Journal
  of Physics A: Mathematical and Theoretical}\ }\textbf {\bibinfo {volume}
  {47}},\ \bibinfo {pages} {424018} (\bibinfo {year} {2014})}\BibitemShut
  {NoStop}%
\bibitem [{\citenamefont {Allen}\ \emph {et~al.}(2017)\citenamefont {Allen},
  \citenamefont {Barrett}, \citenamefont {Horsman}, \citenamefont {Lee},\ and\
  \citenamefont {Spekkens}}]{Allen2016}%
  \BibitemOpen
  \bibfield  {author} {\bibinfo {author} {\bibfnamefont {J.-M.~A.}\
  \bibnamefont {Allen}}, \bibinfo {author} {\bibfnamefont {J.}~\bibnamefont
  {Barrett}}, \bibinfo {author} {\bibfnamefont {D.~C.}\ \bibnamefont
  {Horsman}}, \bibinfo {author} {\bibfnamefont {C.~M.}\ \bibnamefont {Lee}}, \
  and\ \bibinfo {author} {\bibfnamefont {R.~W.}\ \bibnamefont {Spekkens}},\
  }\href@noop {} {\bibfield  {journal} {\bibinfo  {journal} {Physical Review
  X}\ }\textbf {\bibinfo {volume} {7}} (\bibinfo {year} {2017})}\BibitemShut
  {NoStop}%
\bibitem [{\citenamefont {Coretti}\ \emph {et~al.}(2011)\citenamefont
  {Coretti}, \citenamefont {H\"anggi},\ and\ \citenamefont
  {Wolf}}]{Coretti-Hangi-Wolf-2011}%
  \BibitemOpen
  \bibfield  {author} {\bibinfo {author} {\bibfnamefont {S.}~\bibnamefont
  {Coretti}}, \bibinfo {author} {\bibfnamefont {E.}~\bibnamefont {H\"anggi}}, \
  and\ \bibinfo {author} {\bibfnamefont {S.}~\bibnamefont {Wolf}},\ }\href
  {\doibase 10.1103/PhysRevLett.107.100402} {\bibfield  {journal} {\bibinfo
  {journal} {Physical Review Letters}\ }\textbf {\bibinfo {volume} {107}},\
  \bibinfo {pages} {100402} (\bibinfo {year} {2011})}\BibitemShut {NoStop}%
\bibitem [{\citenamefont {Barnea}\ \emph {et~al.}(2013)\citenamefont {Barnea},
  \citenamefont {Bancal}, \citenamefont {Liang},\ and\ \citenamefont
  {Gisin}}]{Barnea-Bancal-et-al:2013}%
  \BibitemOpen
  \bibfield  {author} {\bibinfo {author} {\bibfnamefont {T.~J.}\ \bibnamefont
  {Barnea}}, \bibinfo {author} {\bibfnamefont {J.-D.}\ \bibnamefont {Bancal}},
  \bibinfo {author} {\bibfnamefont {Y.-C.}\ \bibnamefont {Liang}}, \ and\
  \bibinfo {author} {\bibfnamefont {N.}~\bibnamefont {Gisin}},\ }\href
  {\doibase 10.1103/PhysRevA.88.022123} {\bibfield  {journal} {\bibinfo
  {journal} {Physical Review A}\ }\textbf {\bibinfo {volume} {88}},\ \bibinfo
  {pages} {022123} (\bibinfo {year} {2013})}\BibitemShut {NoStop}%
\bibitem [{\citenamefont {Bancal}\ \emph {et~al.}(2012)\citenamefont {Bancal},
  \citenamefont {Pironio}, \citenamefont {Acin}, \citenamefont {Liang},
  \citenamefont {Scarani},\ and\ \citenamefont {Gisin}}]{Bancal2012}%
  \BibitemOpen
  \bibfield  {author} {\bibinfo {author} {\bibfnamefont {J.-D.}\ \bibnamefont
  {Bancal}}, \bibinfo {author} {\bibfnamefont {S.}~\bibnamefont {Pironio}},
  \bibinfo {author} {\bibfnamefont {A.}~\bibnamefont {Acin}}, \bibinfo {author}
  {\bibfnamefont {Y.-C.}\ \bibnamefont {Liang}}, \bibinfo {author}
  {\bibfnamefont {V.}~\bibnamefont {Scarani}}, \ and\ \bibinfo {author}
  {\bibfnamefont {N.}~\bibnamefont {Gisin}},\ }\href {\doibase
  10.1038/nphys2460} {\bibfield  {journal} {\bibinfo  {journal} {Nature
  Physics}\ }\textbf {\bibinfo {volume} {8}},\ \bibinfo {pages} {867} (\bibinfo
  {year} {2012})}\BibitemShut {NoStop}%
\bibitem [{\citenamefont {Stefanov}\ \emph {et~al.}(2002)\citenamefont
  {Stefanov}, \citenamefont {Zbinden}, \citenamefont {Gisin},\ and\
  \citenamefont {Suarez}}]{Stefanov-Zbinden-Gisin-Suarez-2002}%
  \BibitemOpen
  \bibfield  {author} {\bibinfo {author} {\bibfnamefont {A.}~\bibnamefont
  {Stefanov}}, \bibinfo {author} {\bibfnamefont {H.}~\bibnamefont {Zbinden}},
  \bibinfo {author} {\bibfnamefont {N.}~\bibnamefont {Gisin}}, \ and\ \bibinfo
  {author} {\bibfnamefont {A.}~\bibnamefont {Suarez}},\ }\href {\doibase
  10.1103/PhysRevLett.88.120404} {\bibfield  {journal} {\bibinfo  {journal}
  {Physical Review Letters}\ }\textbf {\bibinfo {volume} {88}},\ \bibinfo
  {pages} {120404} (\bibinfo {year} {2002})}\BibitemShut {NoStop}%
\bibitem [{\citenamefont {Wood}\ and\ \citenamefont
  {Spekkens}(2015)}]{finetune}%
  \BibitemOpen
  \bibfield  {author} {\bibinfo {author} {\bibfnamefont {C.~J.}\ \bibnamefont
  {Wood}}\ and\ \bibinfo {author} {\bibfnamefont {R.~W.}\ \bibnamefont
  {Spekkens}},\ }\href {\doibase 10.1088/1367-2630/17/3/033002} {\bibfield
  {journal} {\bibinfo  {journal} {New Journal of Physics}\ }\textbf {\bibinfo
  {volume} {17}},\ \bibinfo {pages} {033002} (\bibinfo {year}
  {2015})}\BibitemShut {NoStop}%
\bibitem [{\citenamefont {Cavalcanti}(2017)}]{Cavalcanti:2017}%
  \BibitemOpen
  \bibfield  {author} {\bibinfo {author} {\bibfnamefont {E.~G.}\ \bibnamefont
  {Cavalcanti}},\ }\href@noop {} {\bibfield  {journal} {\bibinfo  {journal}
  {preprint arXiv:1705.05961 [quant-ph]}\ } (\bibinfo {year}
  {2017})}\BibitemShut {NoStop}%
\bibitem [{\citenamefont {{Costa de Beauregard}}(1953)}]{debeauregard1953}%
  \BibitemOpen
  \bibfield  {author} {\bibinfo {author} {\bibfnamefont {O.}~\bibnamefont
  {{Costa de Beauregard}}},\ }\href@noop {} {\bibfield  {journal} {\bibinfo
  {journal} {Comptes rendus des s\'eances de l'Acad\'emie des Sciences}\
  }\textbf {\bibinfo {volume} {236}},\ \bibinfo {pages} {1632} (\bibinfo {year}
  {1953})}\BibitemShut {NoStop}%
\bibitem [{\citenamefont {Price}(1994)}]{Price1994}%
  \BibitemOpen
  \bibfield  {author} {\bibinfo {author} {\bibfnamefont {H.}~\bibnamefont
  {Price}},\ }\href {\doibase 10.1093/mind/103.411.303} {\bibfield  {journal}
  {\bibinfo  {journal} {Mind}\ }\textbf {\bibinfo {volume} {103}},\ \bibinfo
  {pages} {303} (\bibinfo {year} {1994})}\BibitemShut {NoStop}%
\bibitem [{\citenamefont {Brassard}\ and\ \citenamefont
  {Raymond-Robichaud}(2013)}]{Brassard2013}%
  \BibitemOpen
  \bibfield  {author} {\bibinfo {author} {\bibfnamefont {G.}~\bibnamefont
  {Brassard}}\ and\ \bibinfo {author} {\bibfnamefont {P.}~\bibnamefont
  {Raymond-Robichaud}},\ }in\ \href {\doibase 10.1007/978-1-4614-5212-6_4}
  {\emph {\bibinfo {booktitle} {Is Science Compatible with Free Will?}}},\
  \bibinfo {editor} {edited by\ \bibinfo {editor} {\bibfnamefont
  {A.}~\bibnamefont {Suarez}}\ and\ \bibinfo {editor} {\bibfnamefont
  {P.}~\bibnamefont {Adams}}}\ (\bibinfo  {publisher} {Springer New York},\
  \bibinfo {address} {New York, NY},\ \bibinfo {year} {2013})\ Chap.~\bibinfo
  {chapter} {4}, pp.\ \bibinfo {pages} {41--61}\BibitemShut {NoStop}%
\bibitem [{\citenamefont {Fitzi}(2008)}]{Fitzi08}%
  \BibitemOpen
  \bibfield  {author} {\bibinfo {author} {\bibfnamefont {M.}~\bibnamefont
  {Fitzi}},\ }\href@noop {} {}\bibinfo {howpublished} {Personal communication}
  (\bibinfo {year} {2008})\BibitemShut {NoStop}%
\bibitem [{\citenamefont {Lanczos}(1924)}]{Lanczos:1924}%
  \BibitemOpen
  \bibfield  {author} {\bibinfo {author} {\bibfnamefont {K.}~\bibnamefont
  {Lanczos}},\ }\href {\doibase 10.1007/BF01328251} {\bibfield  {journal}
  {\bibinfo  {journal} {Zeitschrift f{\"u}r Physik}\ }\textbf {\bibinfo
  {volume} {21}},\ \bibinfo {pages} {73} (\bibinfo {year} {1924})}\BibitemShut
  {NoStop}%
\bibitem [{\citenamefont {G{\"{o}}del}(1949)}]{Godel1949}%
  \BibitemOpen
  \bibfield  {author} {\bibinfo {author} {\bibfnamefont {K.}~\bibnamefont
  {G{\"{o}}del}},\ }\href {\doibase 10.1103/RevModPhys.21.447} {\bibfield
  {journal} {\bibinfo  {journal} {Reviews of Modern Physics}\ }\textbf
  {\bibinfo {volume} {21}},\ \bibinfo {pages} {447} (\bibinfo {year}
  {1949})}\BibitemShut {NoStop}%
\bibitem [{\citenamefont {Deutsch}(1991)}]{Deutsch1991}%
  \BibitemOpen
  \bibfield  {author} {\bibinfo {author} {\bibfnamefont {D.}~\bibnamefont
  {Deutsch}},\ }\href {\doibase 10.1103/PhysRevD.44.3197} {\bibfield  {journal}
  {\bibinfo  {journal} {Physical Review D}\ }\textbf {\bibinfo {volume} {44}},\
  \bibinfo {pages} {3197} (\bibinfo {year} {1991})}\BibitemShut {NoStop}%
\bibitem [{\citenamefont {von Neumann}(1932)}]{VonNeumann1932}%
  \BibitemOpen
  \bibfield  {author} {\bibinfo {author} {\bibfnamefont {J.}~\bibnamefont {von
  Neumann}},\ }\href@noop {} {\emph {\bibinfo {title} {Mathematische
  {G}rundlagen der {Q}uantenmechanik}}}\ (\bibinfo  {publisher} {Julius
  Springer},\ \bibinfo {address} {Berlin},\ \bibinfo {year} {1932})\BibitemShut
  {NoStop}%
\bibitem [{\citenamefont {Everett}(1957)}]{Everett1957}%
  \BibitemOpen
  \bibfield  {author} {\bibinfo {author} {\bibfnamefont {H.}~\bibnamefont
  {Everett}},\ }\href {\doibase 10.1103/RevModPhys.29.454} {\bibfield
  {journal} {\bibinfo  {journal} {Reviews of Modern Physics}\ }\textbf
  {\bibinfo {volume} {29}},\ \bibinfo {pages} {454} (\bibinfo {year}
  {1957})}\BibitemShut {NoStop}%
\bibitem [{\citenamefont {Everett}(1973)}]{EverettThesis}%
  \BibitemOpen
  \bibfield  {author} {\bibinfo {author} {\bibfnamefont {H.}~\bibnamefont
  {Everett}},\ }in\ \href@noop {} {\emph {\bibinfo {booktitle} {The Many-Worlds
  Interpretation of Quantum Mechanics}}}\ (\bibinfo  {publisher} {Princeton
  University Press},\ \bibinfo {address} {Princeton},\ \bibinfo {year} {1973})\
  pp.\ \bibinfo {pages} {3--140}\BibitemShut {NoStop}%
\bibitem [{\citenamefont {Henry-Hermann}(1948)}]{Hermann48}%
  \BibitemOpen
  \bibfield  {author} {\bibinfo {author} {\bibfnamefont {G.}~\bibnamefont
  {Henry-Hermann}},\ }\href@noop {} {\bibfield  {journal} {\bibinfo  {journal}
  {Studium Generale}\ }\textbf {\bibinfo {volume} {1}},\ \bibinfo {pages} {375}
  (\bibinfo {year} {1948})}\BibitemShut {NoStop}%
\bibitem [{\citenamefont {Hermann}(1935)}]{HermannFries}%
  \BibitemOpen
  \bibfield  {author} {\bibinfo {author} {\bibfnamefont {G.}~\bibnamefont
  {Hermann}},\ }in\ \href@noop {} {\emph {\bibinfo {booktitle} {Abhand\-lungen
  der Fries'schen Schule. Neue Folge.}}},\ Vol.\ \bibinfo {volume} {6.2},\
  \bibinfo {editor} {edited by\ \bibinfo {editor} {\bibfnamefont
  {O.}~\bibnamefont {Meyerhof}}, \bibinfo {editor} {\bibfnamefont
  {F.}~\bibnamefont {Oppenheimer}}, \ and\ \bibinfo {editor} {\bibfnamefont
  {M.}~\bibnamefont {Specht}}}\ (\bibinfo  {publisher} {Verlag
  ``{\"{O}}ffentliches Leben''},\ \bibinfo {address} {Berlin},\ \bibinfo {year}
  {1935})\ Chap.\ \bibinfo {chapter} {III}, pp.\ \bibinfo {pages}
  {69--152}\BibitemShut {NoStop}%
\bibitem [{\citenamefont {Hermann}(1936)}]{Hermann1936}%
  \BibitemOpen
  \bibfield  {author} {\bibinfo {author} {\bibfnamefont {G.}~\bibnamefont
  {Hermann}},\ }\href@noop {} {\bibfield  {journal} {\bibinfo  {journal}
  {Erkenntnis}\ }\textbf {\bibinfo {volume} {6}},\ \bibinfo {pages} {342}
  (\bibinfo {year} {1936})}\BibitemShut {NoStop}%
\bibitem [{\citenamefont {DeWitt}(1970)}]{DeWitt1970}%
  \BibitemOpen
  \bibfield  {author} {\bibinfo {author} {\bibfnamefont {B.~S.}\ \bibnamefont
  {DeWitt}},\ }\href {\doibase 10.1063/1.3022331} {\bibfield  {journal}
  {\bibinfo  {journal} {Physics Today}\ }\textbf {\bibinfo {volume} {23}},\
  \bibinfo {pages} {30} (\bibinfo {year} {1970})}\BibitemShut {NoStop}%
\bibitem [{\citenamefont {Deutsch}(1997)}]{FabricOfReality}%
  \BibitemOpen
  \bibfield  {author} {\bibinfo {author} {\bibfnamefont {D.}~\bibnamefont
  {Deutsch}},\ }\href@noop {} {\emph {\bibinfo {title} {The Fabric of Reality:
  The Science of Parallel Universes and Its Implications}}}\ (\bibinfo
  {publisher} {Viking Adult},\ \bibinfo {address} {New York},\ \bibinfo {year}
  {1997})\BibitemShut {NoStop}%
\bibitem [{\citenamefont {L{\'{e}}vy-Leblond}(1976)}]{Levy-Leblond1976}%
  \BibitemOpen
  \bibfield  {author} {\bibinfo {author} {\bibfnamefont {J.-M.}\ \bibnamefont
  {L{\'{e}}vy-Leblond}},\ }\href {\doibase 10.1111/j.1746-8361.1976.tb00727.x}
  {\bibfield  {journal} {\bibinfo  {journal} {dialectica}\ }\textbf {\bibinfo
  {volume} {30}},\ \bibinfo {pages} {161} (\bibinfo {year} {1976})}\BibitemShut
  {NoStop}%
\bibitem [{\citenamefont {Page}\ and\ \citenamefont
  {Wootters}(1983)}]{Page1983}%
  \BibitemOpen
  \bibfield  {author} {\bibinfo {author} {\bibfnamefont {D.~N.}\ \bibnamefont
  {Page}}\ and\ \bibinfo {author} {\bibfnamefont {W.~K.}\ \bibnamefont
  {Wootters}},\ }\href {\doibase 10.1103/PhysRevD.27.2885} {\bibfield
  {journal} {\bibinfo  {journal} {Physical Review D}\ }\textbf {\bibinfo
  {volume} {27}},\ \bibinfo {pages} {2885} (\bibinfo {year}
  {1983})}\BibitemShut {NoStop}%
\bibitem [{\citenamefont {Wootters}(1984)}]{Wootters1984}%
  \BibitemOpen
  \bibfield  {author} {\bibinfo {author} {\bibfnamefont {W.~K.}\ \bibnamefont
  {Wootters}},\ }\href {\doibase 10.1007/BF02214098} {\bibfield  {journal}
  {\bibinfo  {journal} {International Journal of Theoretical Physics}\ }\textbf
  {\bibinfo {volume} {23}},\ \bibinfo {pages} {701} (\bibinfo {year}
  {1984})}\BibitemShut {NoStop}%
\bibitem [{\citenamefont {{Costa de Beauregard}}()}]{debeauregard-interview}%
  \BibitemOpen
  \bibfield  {author} {\bibinfo {author} {\bibfnamefont {O.}~\bibnamefont
  {{Costa de Beauregard}}},\ }\href@noop {} {}\bibinfo {howpublished}
  {{Interview with Olivier Costa de Beauregard by Solange Collery. Tonus, 2
  Novembre 1981}}\BibitemShut {NoStop}%
\bibitem [{\citenamefont {{Costa de Beauregard}}(1976)}]{DeBeauregard1976}%
  \BibitemOpen
  \bibfield  {author} {\bibinfo {author} {\bibfnamefont {O.}~\bibnamefont
  {{Costa de Beauregard}}},\ }\href {\doibase 10.1007/BF00715107} {\bibfield
  {journal} {\bibinfo  {journal} {Foundations of Physics}\ }\textbf {\bibinfo
  {volume} {6}},\ \bibinfo {pages} {539} (\bibinfo {year} {1976})}\BibitemShut
  {NoStop}%
\bibitem [{\citenamefont {{Costa de Beauregard}}(1977)}]{DeBeauregard1977}%
  \BibitemOpen
  \bibfield  {author} {\bibinfo {author} {\bibfnamefont {O.}~\bibnamefont
  {{Costa de Beauregard}}},\ }\href {\doibase 10.1007/BF02906749} {\bibfield
  {journal} {\bibinfo  {journal} {Il Nuovo Cimento B}\ }\textbf {\bibinfo
  {volume} {42}},\ \bibinfo {pages} {41} (\bibinfo {year} {1977})}\BibitemShut
  {NoStop}%
\bibitem [{\citenamefont {Bennett}\ and\ \citenamefont
  {Wiesner}(1992)}]{Bennett-super-dense-coding:1992}%
  \BibitemOpen
  \bibfield  {author} {\bibinfo {author} {\bibfnamefont {C.~H.}\ \bibnamefont
  {Bennett}}\ and\ \bibinfo {author} {\bibfnamefont {S.~J.}\ \bibnamefont
  {Wiesner}},\ }\href {\doibase 10.1103/PhysRevLett.69.2881} {\bibfield
  {journal} {\bibinfo  {journal} {Physical Review Letters}\ }\textbf {\bibinfo
  {volume} {69}},\ \bibinfo {pages} {2881} (\bibinfo {year}
  {1992})}\BibitemShut {NoStop}%
\bibitem [{\citenamefont {Feldmann}(1995)}]{feldmann95}%
  \BibitemOpen
  \bibfield  {author} {\bibinfo {author} {\bibfnamefont {M.}~\bibnamefont
  {Feldmann}},\ }\href@noop {} {\bibfield  {journal} {\bibinfo  {journal}
  {Foundations of Physics Letters}\ }\textbf {\bibinfo {volume} {8}},\ \bibinfo
  {pages} {41} (\bibinfo {year} {1995})}\BibitemShut {NoStop}%
\bibitem [{\citenamefont {Brans}(1988)}]{Brans1988}%
  \BibitemOpen
  \bibfield  {author} {\bibinfo {author} {\bibfnamefont {C.~H.}\ \bibnamefont
  {Brans}},\ }\href {\doibase 10.1007/BF00670750} {\bibfield  {journal}
  {\bibinfo  {journal} {International Journal of Theoretical Physics}\ }\textbf
  {\bibinfo {volume} {27}},\ \bibinfo {pages} {219} (\bibinfo {year}
  {1988})}\BibitemShut {NoStop}%
\bibitem [{\citenamefont {Hall}(2016)}]{Hall2016}%
  \BibitemOpen
  \bibfield  {author} {\bibinfo {author} {\bibfnamefont {M.~J.~W.}\
  \bibnamefont {Hall}},\ }in\ \href {\doibase 10.1007/978-3-319-31299-6_11}
  {\emph {\bibinfo {booktitle} {At the Frontier of Spacetime: Scalar-Tensor
  Theory, {B}ell's Inequality, {M}ach's Principle, Exotic Smoothness}}},\
  \bibinfo {editor} {edited by\ \bibinfo {editor} {\bibfnamefont
  {T.}~\bibnamefont {Asselmeyer-Maluga}}}\ (\bibinfo  {publisher} {Springer
  International Publishing},\ \bibinfo {address} {Cham},\ \bibinfo {year}
  {2016})\ pp.\ \bibinfo {pages} {189--204}\BibitemShut {NoStop}%
\bibitem [{\citenamefont {Degorre}\ \emph {et~al.}(2005)\citenamefont
  {Degorre}, \citenamefont {Laplante},\ and\ \citenamefont
  {Roland}}]{Degorre2005}%
  \BibitemOpen
  \bibfield  {author} {\bibinfo {author} {\bibfnamefont {J.}~\bibnamefont
  {Degorre}}, \bibinfo {author} {\bibfnamefont {S.}~\bibnamefont {Laplante}}, \
  and\ \bibinfo {author} {\bibfnamefont {J.}~\bibnamefont {Roland}},\ }\href
  {\doibase 10.1103/PhysRevA.72.062314} {\bibfield  {journal} {\bibinfo
  {journal} {Physical Review A}\ }\textbf {\bibinfo {volume} {72}},\ \bibinfo
  {pages} {062314} (\bibinfo {year} {2005})}\BibitemShut {NoStop}%
\bibitem [{\citenamefont {Toner}\ and\ \citenamefont
  {Bacon}(2003)}]{Toner:2003dp}%
  \BibitemOpen
  \bibfield  {author} {\bibinfo {author} {\bibfnamefont {B.~F.}\ \bibnamefont
  {Toner}}\ and\ \bibinfo {author} {\bibfnamefont {D.}~\bibnamefont {Bacon}},\
  }\href {\doibase 10.1103/PhysRevLett.91.187904} {\bibfield  {journal}
  {\bibinfo  {journal} {Physical Review Letters}\ }\textbf {\bibinfo {volume}
  {91}},\ \bibinfo {pages} {187904} (\bibinfo {year} {2003})}\BibitemShut
  {NoStop}%
\bibitem [{\citenamefont {Kofler}\ \emph {et~al.}(2006)\citenamefont {Kofler},
  \citenamefont {Paterek},\ and\ \citenamefont
  {Brukner}}]{Kofler-Paterek-Brukner:2006}%
  \BibitemOpen
  \bibfield  {author} {\bibinfo {author} {\bibfnamefont {J.}~\bibnamefont
  {Kofler}}, \bibinfo {author} {\bibfnamefont {T.}~\bibnamefont {Paterek}}, \
  and\ \bibinfo {author} {\bibfnamefont {{\v{C}}.}~\bibnamefont {Brukner}},\
  }\href {\doibase 10.1103/PhysRevA.73.022104} {\bibfield  {journal} {\bibinfo
  {journal} {Physical Review A}\ }\textbf {\bibinfo {volume} {73}},\ \bibinfo
  {pages} {022104} (\bibinfo {year} {2006})}\BibitemShut {NoStop}%
\bibitem [{\citenamefont {Roland}\ and\ \citenamefont
  {Szegedy}(2009)}]{Roland2009}%
  \BibitemOpen
  \bibfield  {author} {\bibinfo {author} {\bibfnamefont {J.}~\bibnamefont
  {Roland}}\ and\ \bibinfo {author} {\bibfnamefont {M.}~\bibnamefont
  {Szegedy}},\ }\enquote {\bibinfo {title} {Amortized communication complexity
  of distributions},}\ in\ \href {\doibase 10.1007/978-3-642-02927-1_61} {\emph
  {\bibinfo {booktitle} {Automata, Languages and Programming: 36th
  International Colloquium, ICALP 2009, Rhodes, Greece, July 5-12, 2009,
  Proceedings, Part I}}},\ \bibinfo {editor} {edited by\ \bibinfo {editor}
  {\bibfnamefont {S.}~\bibnamefont {Albers}}, \bibinfo {editor} {\bibfnamefont
  {A.}~\bibnamefont {Marchetti-Spaccamela}}, \bibinfo {editor} {\bibfnamefont
  {Y.}~\bibnamefont {Matias}}, \bibinfo {editor} {\bibfnamefont
  {S.}~\bibnamefont {Nikoletseas}}, \ and\ \bibinfo {editor} {\bibfnamefont
  {W.}~\bibnamefont {Thomas}}}\ (\bibinfo  {publisher} {Springer Berlin
  Heidelberg},\ \bibinfo {address} {Berlin, Heidelberg},\ \bibinfo {year}
  {2009})\ pp.\ \bibinfo {pages} {738--749}\BibitemShut {NoStop}%
\bibitem [{\citenamefont {Barrett}\ and\ \citenamefont
  {Gisin}(2011)}]{Gisin-Barett:2011}%
  \BibitemOpen
  \bibfield  {author} {\bibinfo {author} {\bibfnamefont {J.}~\bibnamefont
  {Barrett}}\ and\ \bibinfo {author} {\bibfnamefont {N.}~\bibnamefont
  {Gisin}},\ }\href {\doibase 10.1103/PhysRevLett.106.100406} {\bibfield
  {journal} {\bibinfo  {journal} {Physical Review Letters}\ }\textbf {\bibinfo
  {volume} {106}},\ \bibinfo {pages} {100406} (\bibinfo {year}
  {2011})}\BibitemShut {NoStop}%
\bibitem [{\citenamefont {Cerf}\ \emph {et~al.}(2005)\citenamefont {Cerf},
  \citenamefont {Gisin}, \citenamefont {Massar},\ and\ \citenamefont
  {Popescu}}]{cgmp05}%
  \BibitemOpen
  \bibfield  {author} {\bibinfo {author} {\bibfnamefont {N.~J.}\ \bibnamefont
  {Cerf}}, \bibinfo {author} {\bibfnamefont {N.}~\bibnamefont {Gisin}},
  \bibinfo {author} {\bibfnamefont {S.}~\bibnamefont {Massar}}, \ and\ \bibinfo
  {author} {\bibfnamefont {S.}~\bibnamefont {Popescu}},\ }\href@noop {}
  {\bibfield  {journal} {\bibinfo  {journal} {Physical Review Letters}\
  }\textbf {\bibinfo {volume} {94}},\ \bibinfo {pages} {220403} (\bibinfo
  {year} {2005})},\ \bibinfo {note} {quant-ph/0410027}\BibitemShut {NoStop}%
\bibitem [{\citenamefont {Aaronson}\ and\ \citenamefont
  {Watrous}(2009)}]{Aaronson:2009}%
  \BibitemOpen
  \bibfield  {author} {\bibinfo {author} {\bibfnamefont {S.}~\bibnamefont
  {Aaronson}}\ and\ \bibinfo {author} {\bibfnamefont {J.}~\bibnamefont
  {Watrous}},\ }\href {\doibase 10.1098/rspa.2008.0350} {\bibfield  {journal}
  {\bibinfo  {journal} {Proceedings of the Royal Society A: Mathematical,
  Physical and Engineering Sciences}\ }\textbf {\bibinfo {volume} {465}},\
  \bibinfo {pages} {631} (\bibinfo {year} {2009})}\BibitemShut {NoStop}%
\bibitem [{\citenamefont {Bacon}(2004)}]{Bacon-ctc:2004}%
  \BibitemOpen
  \bibfield  {author} {\bibinfo {author} {\bibfnamefont {D.}~\bibnamefont
  {Bacon}},\ }\href {\doibase 10.1103/PhysRevA.70.032309} {\bibfield  {journal}
  {\bibinfo  {journal} {Physical Review A}\ }\textbf {\bibinfo {volume} {70}},\
  \bibinfo {pages} {032309} (\bibinfo {year} {2004})}\BibitemShut {NoStop}%
\bibitem [{\citenamefont {Yuan}\ \emph {et~al.}(2015)\citenamefont {Yuan},
  \citenamefont {Assad}, \citenamefont {Thompson}, \citenamefont {Haw},
  \citenamefont {Vedral}, \citenamefont {Ralph}, \citenamefont {Lam},
  \citenamefont {Weedbrook},\ and\ \citenamefont {Gu}}]{Mile2015}%
  \BibitemOpen
  \bibfield  {author} {\bibinfo {author} {\bibfnamefont {X.}~\bibnamefont
  {Yuan}}, \bibinfo {author} {\bibfnamefont {S.~M.}\ \bibnamefont {Assad}},
  \bibinfo {author} {\bibfnamefont {J.}~\bibnamefont {Thompson}}, \bibinfo
  {author} {\bibfnamefont {J.~Y.}\ \bibnamefont {Haw}}, \bibinfo {author}
  {\bibfnamefont {V.}~\bibnamefont {Vedral}}, \bibinfo {author} {\bibfnamefont
  {T.~C.}\ \bibnamefont {Ralph}}, \bibinfo {author} {\bibfnamefont {P.~K.}\
  \bibnamefont {Lam}}, \bibinfo {author} {\bibfnamefont {C.}~\bibnamefont
  {Weedbrook}}, \ and\ \bibinfo {author} {\bibfnamefont {M.}~\bibnamefont
  {Gu}},\ }\href {\doibase 10.1038/npjqi.2015.7} {\bibfield  {journal}
  {\bibinfo  {journal} {NPJ Quantum Information}\ }\textbf {\bibinfo {volume}
  {1}},\ \bibinfo {pages} {15007} (\bibinfo {year} {2015})}\BibitemShut
  {NoStop}%
\bibitem [{\citenamefont {Cirel'son}(1980)}]{TsirelsonsBound}%
  \BibitemOpen
  \bibfield  {author} {\bibinfo {author} {\bibfnamefont {B.~S.}\ \bibnamefont
  {Cirel'son}},\ }\href {\doibase 10.1007/BF00417500} {\bibfield  {journal}
  {\bibinfo  {journal} {Letters in Mathematical Physics}\ }\textbf {\bibinfo
  {volume} {4}},\ \bibinfo {pages} {93} (\bibinfo {year} {1980})}\BibitemShut
  {NoStop}%
\bibitem [{\citenamefont {Pienaar}\ \emph {et~al.}(2013)\citenamefont
  {Pienaar}, \citenamefont {Ralph},\ and\ \citenamefont {Myers}}]{Pienaar2013}%
  \BibitemOpen
  \bibfield  {author} {\bibinfo {author} {\bibfnamefont {J.}~\bibnamefont
  {Pienaar}}, \bibinfo {author} {\bibfnamefont {T.~C.}\ \bibnamefont {Ralph}},
  \ and\ \bibinfo {author} {\bibfnamefont {C.~R.}\ \bibnamefont {Myers}},\
  }\href {\doibase 10.1103/PhysRevLett.110.060501} {\bibfield  {journal}
  {\bibinfo  {journal} {Physical Review Letters}\ }\textbf {\bibinfo {volume}
  {110}},\ \bibinfo {pages} {60501} (\bibinfo {year} {2013})}\BibitemShut
  {NoStop}%
\bibitem [{\citenamefont {Deutsch}\ and\ \citenamefont
  {Hayden}(2000)}]{Deutsch-Hayden:2000}%
  \BibitemOpen
  \bibfield  {author} {\bibinfo {author} {\bibfnamefont {D.}~\bibnamefont
  {Deutsch}}\ and\ \bibinfo {author} {\bibfnamefont {P.}~\bibnamefont
  {Hayden}},\ }\href {\doibase 10.1098/rspa.2000.0585} {\bibfield  {journal}
  {\bibinfo  {journal} {Proceedings of the Royal Society of London A:
  Mathematical, Physical and Engineering Sciences}\ }\textbf {\bibinfo {volume}
  {456}},\ \bibinfo {pages} {1759} (\bibinfo {year} {2000})}\BibitemShut
  {NoStop}%
\bibitem [{\citenamefont {Price}\ and\ \citenamefont
  {Wharton}(2015)}]{price2015}%
  \BibitemOpen
  \bibfield  {author} {\bibinfo {author} {\bibfnamefont {H.}~\bibnamefont
  {Price}}\ and\ \bibinfo {author} {\bibfnamefont {K.}~\bibnamefont
  {Wharton}},\ }\href {\doibase 10.3390/e17117752} {\bibfield  {journal}
  {\bibinfo  {journal} {Entropy}\ }\textbf {\bibinfo {volume} {17}},\ \bibinfo
  {pages} {7752} (\bibinfo {year} {2015})}\BibitemShut {NoStop}%
\bibitem [{\citenamefont {Oreshkov}\ \emph {et~al.}(2012)\citenamefont
  {Oreshkov}, \citenamefont {Costa},\ and\ \citenamefont {Brukner}}]{OCB}%
  \BibitemOpen
  \bibfield  {author} {\bibinfo {author} {\bibfnamefont {O.}~\bibnamefont
  {Oreshkov}}, \bibinfo {author} {\bibfnamefont {F.}~\bibnamefont {Costa}}, \
  and\ \bibinfo {author} {\bibfnamefont {{\v{C}}.}~\bibnamefont {Brukner}},\
  }\href {\doibase 10.1038/ncomms2076} {\bibfield  {journal} {\bibinfo
  {journal} {Nature Communications}\ }\textbf {\bibinfo {volume} {3}},\
  \bibinfo {pages} {1092} (\bibinfo {year} {2012})}\BibitemShut {NoStop}%
\bibitem [{\citenamefont {Baumeler}\ and\ \citenamefont
  {Wolf}(2016{\natexlab{a}})}]{Baumeler2016}%
  \BibitemOpen
  \bibfield  {author} {\bibinfo {author} {\bibfnamefont {{\"{A}}.}~\bibnamefont
  {Baumeler}}\ and\ \bibinfo {author} {\bibfnamefont {S.}~\bibnamefont
  {Wolf}},\ }\href {\doibase 10.1088/1367-2630/18/3/035014} {\bibfield
  {journal} {\bibinfo  {journal} {New Journal of Physics}\ }\textbf {\bibinfo
  {volume} {18}},\ \bibinfo {pages} {035014} (\bibinfo {year}
  {2016}{\natexlab{a}})}\BibitemShut {NoStop}%
\bibitem [{\citenamefont {Wheeler}(1990)}]{Wheeler:1989}%
  \BibitemOpen
  \bibfield  {author} {\bibinfo {author} {\bibfnamefont {J.~A.}\ \bibnamefont
  {Wheeler}},\ }in\ \href@noop {} {\emph {\bibinfo {booktitle} {Complexity,
  Entropy, and the Physics of Information: The Proceedings of the Workshop Held
  May-June, 1989, in Santa Fe, New Mexico}}},\ \bibinfo {editor} {edited by\
  \bibinfo {editor} {\bibfnamefont {W.}~\bibnamefont {Zurek}}}\ (\bibinfo
  {publisher} {Avalon Publishing},\ \bibinfo {year} {1990})\ pp.\ \bibinfo
  {pages} {3--28}\BibitemShut {NoStop}%
\bibitem [{\citenamefont {Baumeler}\ and\ \citenamefont
  {Wolf}(2016{\natexlab{b}})}]{Baumeler-Wolf-ccc:2016}%
  \BibitemOpen
  \bibfield  {author} {\bibinfo {author} {\bibfnamefont {{\"{A}}.}~\bibnamefont
  {Baumeler}}\ and\ \bibinfo {author} {\bibfnamefont {S.}~\bibnamefont
  {Wolf}},\ }\href@noop {} {\bibfield  {journal} {\bibinfo  {journal} {preprint
  arXiv:1602.06987v2 [quant-ph]}\ } (\bibinfo {year}
  {2016}{\natexlab{b}})}\BibitemShut {NoStop}%
\end{thebibliography}%

\end{document}